\UseRawInputEncoding 

\documentclass[twocolumn,showpacs,floats,floatfix,superscriptaddress,aps,pra]{revtex4-1}

\usepackage{amsfonts}
\usepackage{amssymb}
\usepackage{amsmath}
\usepackage{calc}
\usepackage{graphicx}
\usepackage{bm}

\usepackage[normalem]{ulem}
\usepackage{amsmath,amssymb}
\usepackage{multirow}
\usepackage{xcolor,soul}
\usepackage{srcltx}
\usepackage{hyperref,graphicx}

\def\be{ \begin{equation}}
\def\ee{ \end{equation}}
\def\bea{ \begin{eqnarray}}
\def\eea{ \end{eqnarray}}
\def\bse{ \begin{subequations}}
\def\ese{ \end{subequations}}
\def\bc{ \begin{center}}
\def\ec{ \end{center}}

\def\Im{\text{Im}\,}

\begin{document}

\author{Stefano Longhi$^{*}$}
\affiliation{Dipartimento di Fisica, Politecnico di Milano, Piazza L. da Vinci 32, I-20133 Milano, Italy}
\affiliation{IFISC (UIB-CSIC), Instituto de Fisica Interdisciplinar y Sistemas Complejos, E-07122 Palma de Mallorca, Spain}
\email{stefano.longhi@polimi.it}

\author{Liang Feng} 
\affiliation{Department of Materials Science and Engineering, University of Pennsylvania, Philadelphia, PA 19104, USA}

\title{Complex Berry phase and imperfect non-Hermitian phase transitions}
  \normalsize


%
\bigskip
\begin{abstract}
\noindent  
In many classical and quantum systems described by an effective non-Hermitian Hamiltonian, spectral phase transitions, from an entirely real energy spectrum to a complex spectrum, can be observed as a non-Hermitian parameter in the system is increased above a critical value. A paradigmatic example is provided by systems possessing parity-time ($\mathcal{PT}$) symmetry, where the energy spectrum remains entirely real in the unbroken $\mathcal{PT}$ phase while a transition to complex energies is observed in the unbroken $\mathcal{PT}$ phase. Such spectral phase transitions are universally sharp. However, when the system is slowly and periodically cycled, the phase transition can become smooth, i.e. imperfect, owing to the complex Berry phase associated to the cyclic adiabatic evolution of the system. This remarkable phenomenon is illustrated by considering the spectral phase transition of the Wannier-Stark ladders in a $\mathcal{PT}$-symmetric class of two-band non-Hermitian lattices subjected to an external dc field, unraveling that a non-vanishing imaginary part of the Zak phase - the Berry phase picked up by a Bloch eigenstate evolving across the entire Brillouin zone- is responsible for imperfect spectral phase transitions.
 \end{abstract}

\maketitle

\section{Introduction}
The geometric or Berry phase \cite{b1,b2,b3,b3bis}, a concept which was systematized and popularized in the 1980s by Sir
Michael Berry \cite{b1}, has permeated through all
branches of physics with applications in diverse fields ranging from atomic and molecular physics \cite{b4,b5,b6} to
condensed-matter physics \cite{b7,b8,b9},  classical optics \cite{b10,b11,b12}, high energy and particle physics \cite{b13,b14,b15}, gravity and cosmology \cite{b16}. 
When a quantum or classical system undergoes a cyclic evolution governed by a change of parameters, besides the dynamical phase it 
 acquires an additional phase term, the Berry phase, which depends only on the geometry of the path but not on how the cycle is run.  
  In condensed matter physics, the geometric phase manifests itself in many phenomena, such as the quantum Hall effect, electric polarization, orbital
 magnetism and exchange statistics \cite{b3,b8,b9}. In a crystal, the application of an electric field changes the
 quasi-momentum of the electronic wave function over the entire Brillouin zone, and the accumulated geometric phase is known as the Zak phase
  \cite{b17}. In one-dimensional (1D) lattices, the bulk topological
properties of the Bloch bands are characterized by the quantized
Zak phase \cite{b18,b19,b20,b21,b22,b23,b23bis}, which can serve as a topological number.\par
Several exciting phenomena that are attracting great interest in modern condensed-matter physics and beyond, such as non-Hermitian skin effect, modified bulk-boundary
correspondence, exceptional points, nontrivial spectral topology and phase transitions, etc. \cite{b24,b25,b26,b27,b27bis},  appear in non-Hermitian models, i.e. in models where the dynamics is described by an effective non-Hermitian 
Hamiltonian \cite{b28,b29,b30} which accounts for energy/particle exchange with external reservoirs. A remarkable property of certain classes of non-Hermitian Hamiltonians is to display an entirely real energy spectrum in spite of non-Hermiticity \cite{b31,b32,b33,b34,b35,b36,b37,b38,b39,b39bis,b40}. Among such Hamiltonians, great attention has been devoted to the ones displaying parity-time ($\mathcal{PT}$) symmetry \cite{b31,b32,b33}, a concept that has become very popular in the past decade and found important applications in photonics and beyond \cite{b41,b42,b43,b44,b45,b46,b47}. 
{\color{black}{For given parity $\mathcal{P}$ and time reversal $\mathcal{T}$ operators, an Hamiltonian $\mathcal{H}$ is said to be $\mathcal{PT}$-symmetric if the commutator $ [\mathcal{H},\mathcal{PT]}  $ vanishes, i.e. $ \mathcal{HPT}=\mathcal{PTH}$. However, since the operator $\mathcal{PT}$ is not linear, 
$\mathcal{PT}$-symmetry itself does not necessarily imply that the $\mathcal{H}$ and $\mathcal{PT}$ operators share the same set of eigenfunctions. {\color{black} This means that, while the underlying Hamiltonian $\mathcal{H}$ possesses $\mathcal{PT}$ symmetry, i.e. $\mathcal{PTH}=\mathcal{HPT}$, the corresponding eigenfunctions $|E \rangle$ of $\mathcal{H}$  can (or cannot) display the same symmetry. When some eigenfunctions of $\mathcal{H}$ break the $\mathcal{PT}$ symmetry, i.e. $\mathcal{PT}|E\rangle$  and $|E \rangle$ are distinct states, we have a typical scenario of spontaneous symmetry breaking} \cite{b33}. Spontaneous $\mathcal{PT}$ symmetry breaking corresponds to a spectral phase transition, from an entirely real energy spectrum in the unbroken $\mathcal{PT}$ phase to a complex energy spectrum in the spontaneously broken $\mathcal{PT}$ phase. {\color{black} {When a control parameter in the system is varied above a critical value, spontaneous $\mathcal{PT}$ symmetry breaking is usually observed and in the broken $\mathcal{PT}$ phase energies appear in complex conjugate pairs. This readily follows from the anti-linear nature of the $\mathcal{T}$ operator}}: {\color{black} if $|E \rangle$ is an eigenfunction of $\mathcal{H}$ with eigenenergy $E$, i.e. $\mathcal{H} |E \rangle= E |E \rangle$, then 
$\mathcal{HPT} |E \rangle=\mathcal{PT H} | E \rangle=\mathcal{PT} E |E \rangle=E^* \mathcal{PT} |E \rangle$. This means that $\mathcal{PT} |E \rangle$ is an eigenfunction of $\mathcal{H}$ with eigenenergy $E^*$. When the symmetry is not spontaneously broken, $| E \rangle$ and $\mathcal{PT} |E \rangle$ are the same eigenfunction, which necessarily implies $E=E^*$: in the unbroken $\mathcal{PT}$ phase the energy spectrum is entirely real. On the other hand, when the symmetry is spontaneously broken, $|E \rangle$ is not necessarily an eigenfunction of the $\mathcal{PT}$ operator, and thus the eigenfunctions $|E \rangle$ and $\mathcal{PT} |E \rangle$, with non-degenerate eigenenergies $E$ and $E^*$, are linearly independent: in this case the energy spectrum becomes complex and formed by complex conjugate pairs.}
{\color{black}{ The spontaneous symmetry breaking phase transition is ubiquitously sharp and the symmetry breaking point corresponds to the appearance of non-Hermitian degeneracies, i.e. exceptional points \cite{b47,b47bis,b59} or spectral singularities  \cite{sp1,sp2,sp3}, at the critical point.}}\\
The concept of geometric phase can be generalized
to non-Hermitian systems, providing a geometrical description
of the quantum evolution of non-Hermitian systems under cyclic variation of parameters \cite{b48,b49,b50,b51,b52,b53,b54,new1,b54noo,b54basta1,b54basta2,b54bis,b55,b56,b55bis,b57}. As compared to Hermitian systems, different forms of Berry phases have been introduced. Here we will use the Berry phase  from the biorthogonal basis of the non-Hermitian Hamiltonain, which is thus rather generally complex. 
An interesting property of adiabatic cycling in non-Hermitian systems is that the energy surface can display a nontrivial topology: when one follows a loop in the space of system parameters, even in the absence of degeneracies the energies and corresponding instantaneous eigenstates may swap
places, which renders the evolution non-cyclic. The interchange of
energies arises when exceptional points are encircled in the space of system parameters \cite{b47,b58,b59}. The complex Berry phase has been 
suggested to provide a topological invariant identifying different topological phases and quantum phase transitions in certain non-Hermitian  models \cite{b27bis,b60,b61,b62,b62bis,b63,b63bis,b64,b65,b66,b66bis,b66tris,b67,b68,b69,b70}, and some general conditions for the quantization of the Berry phase under certain generalized symmetries have been provided \cite{b71}. In a non-Hermitian lattice,
complex Berry phase, i.e. Zak phase, naturally arises under an external dc force or a time-varying magnetic flux \cite{b65}, so that a Bloch eigenstate adiabatically evolves across the entire Brillouin zone accumulating a complex geometric phase. While the related phenomena of Bloch oscillations and Zener tunneling have been investigated at some extent in non-Hermitian lattices \cite{b72,b72bis,b73,b74,b75,b76,b77,b78,b79,b80,b81}, physical signatures of the complex Zak phase have received so far little attention and mostly restricted to some specific lattice models \cite{b54basta1,b65}.  \par
In this work we show that the complex Berry phase in slow-cycled non-Hermitian $\mathcal{PT}$ symmetric systems can lead to imperfect, i.e. smooth, spectral phase transitions. This phenomenon is first illustrated by considering a general model of two-level $\mathcal{PT}$ symmetric systems, and then applied to explain the imperfect phase transition of Wannier-Stark ladders found in certain two-band non-Hermitian lattices  \cite{b81}, which is rooted in the non-vanishing imaginary part of the Zak phase.

\section{Phase transitions in a cycled two-level $\mathcal{PT}$ symmetric model}
\subsection{Model and $\mathcal{PT}$ symmetry breaking phase transition}
We consider  a classical or quantum two-level system described by an effective $2 \times 2$ non-Hermitian matrix Hamiltonian $\mathcal{H}=\mathcal{H}(k)$, which depends on a real parameter $k$ and is periodic in $k$ with a period of $ 2 \pi$, i.e. $\mathcal{H}(k+2\pi)=\mathcal{H}(k)$. As we will discuss in the next section, in the Wannier-Stark ladder problem of a non-Hermitian lattice driven by a dc field the matrix Hamiltonian $\mathcal{H}(k)$ corresponds to the Bloch Hamiltonian of a two-band lattice and $k$ is the quasi momentum, that drifts to span the entire Brillouin zone in the presence of the dc field.\\
The temporal dynamics of the system is described by the Schr\"odinger equation
\begin{equation}
i \frac{d}{dt} 
\left(
\begin{array}{c}
\psi_1 \\
\psi_2
\end{array}
\right)
= \left(
\begin{array}{cc}
\mathcal{H}_{11} & \mathcal{H}_{12} \\
\mathcal{H}_{21} & \mathcal{H}_{22}
\end{array}
\right)
 \left(
\begin{array}{c}
\psi_1 \\
\psi_2
\end{array}
\right)
=
\mathcal{H}(k)  \left(
\begin{array}{c}
\psi_1 \\
\psi_2
\end{array}
\right).
\end{equation}
 We assume that the Hamiltonian is $\mathcal{PT}$ symmetric with parity $\mathcal{P}$ and time reversal $\mathcal{T}$ operators defined by
 \begin{equation}
 \mathcal{P}= \sigma_x=
 \left(
 \begin{array}{cc}
0 & 1 \\
1 & 0
\end{array}
\right) \; ,\;\; \mathcal{T}=\mathcal{K}
 \end{equation} 
where $\sigma_x$ is the Pauli matrix and $\mathcal{K}$ the element-wise complex conjugation operator. $\mathcal{PT}$ symmetry, i.e. the condition $\mathcal{PTH}=\mathcal{H PT}$, is satisfied provided that
\begin{equation}
\mathcal{H}_{22}= \mathcal{H}^*_{11} \; , \; \; \mathcal{H}_{21}= \mathcal{H}^*_{12}
\end{equation}
so that the non-Hermiticity in the system is embedded in a non-vanishing imaginary part of $\mathcal{H}_{11}$.
The most general form of matrix elements that respect the $\mathcal{PT}$ symmetry is thus
\begin{eqnarray}
\mathcal{H}_{11}=\mathcal{H}_{22}^*=G(k)+i \lambda W(k) \\
 \mathcal{H}_{12}=\mathcal{H}_{21}^{*}=R(k) \exp [i \varphi(k)]
\end{eqnarray}
where $G(k), W(k)$, $R(k)$ are real and periodic functions of $k$ with period $2 \pi$, $\varphi(k)$ is a real function with $\varphi(k+2 \pi)=\varphi(k)$ mod $ 2 \pi$, and $\lambda \geq 0$ is a real parameter 
that measures the strength of non-Hermiticity in the system, the case $\lambda=0$ corresponding to $\mathcal{H}(k)$ Hermitian. Further, we assume that $R(k)$ is nonvanishing over the entire range $0 \leq k \leq 2 \pi$.\\
When the parameter $k$ is kept constant, the eigenenergies of $\mathcal{H}(k)$ are given by
\begin{equation}
E_{\pm}(k)= G(k) \pm \sqrt{R^2(k)-\lambda^2 W^2(k)}
\end{equation}
with corresponding (right) eigenvectors 
\begin{eqnarray}
\mathbf{u}_{+} (k) & = &  \left(
\begin{array}{c}
\cos \left( \frac{\theta}{2} \right) \\
\sin \left( \frac{\theta}{2} \right) \exp(-i \varphi)
\end{array}
\right) \\
 \mathbf{u}_{-} (k) & = &  \left(
\begin{array}{c}
\sin \left( \frac{\theta}{2} \right) \\
-\cos \left( \frac{\theta}{2} \right) \exp(-i \varphi)
\end{array}
\right).
\end{eqnarray}
In the previous equations, the complex angle $\theta=\theta(k)$ is defined by the relation
\begin{equation}
\tan \theta(k) = \frac{R(k)}{i \lambda W(k)}.
\end{equation}
Note that the imaginary part of the angle $\theta(k)$ diverges when $R(k)= \pm \lambda W(k)$, corresponding to the simultaneous coalescence of the two energies and eigenstates, i.e. to the appearance of an exceptional point. 
The left eigenvectors of $\mathcal{H}(k)$, i.e. the (right) eigenvectors of the adjoint $\mathcal{H}^{\dag}(k)$ with eigenvalues $ E_{\pm}^*(k)$, read
\begin{eqnarray}
\mathbf{v}_{+} (k) & = &  \left(
\begin{array}{c}
\cos^* \left( \frac{\theta}{2} \right) \\
\sin^* \left( \frac{\theta}{2} \right) \exp(-i \varphi)
\end{array}
\right) \\
  \mathbf{v}_{-} (k) & = &  \left(
\begin{array}{c}
\sin^* \left( \frac{\theta}{2} \right) \\
-\cos^* \left( \frac{\theta}{2} \right) \exp(-i \varphi)
\end{array}
\right)
\end{eqnarray}
and the biorthogonal conditions
\begin{equation}
\langle \mathbf{v}_{n} (k) | \mathbf{u}_{m} (k) \rangle = \delta_{n,m}
\end{equation}
are satisfied for any $k$, with $n,m=+,-$. 
After letting 
\begin{equation}
\lambda_c(k) \equiv |R(k)/W(k)|,
\end{equation}
 from Eq.(6) it readily follows that the energy spectrum is real for $\lambda < \lambda_c(k)$ (unbroken $\mathcal{PT}$ phase), and complex for $\lambda> \lambda_c(k)$ (broken $\mathcal{PT}$ phase), with the appearance of an exceptional point at the critical point $\lambda=\lambda_c(k)$.

\subsection{Phase transition in the cycled system}
Let us now consider the two-level system when the Hamiltonian $\mathcal{H}(k)$ is periodically and adiabatically cycled in time. We assume that the parameter $k$ in the Hamiltonian varies in time according to
\begin{equation}
k= \omega t
\end{equation}
where $\omega$ is the cycling frequency. The temporal dynamics of the two-level system is thus described by the equation
\begin{equation}
i \frac{d}{dt} 
\left(
\begin{array}{c}
\psi_1 \\
\psi_2
\end{array}
\right)
=
\mathcal{H}( \omega t )  \left(
\begin{array}{c}
\psi_1 \\
\psi_2
\end{array}
\right).
\end{equation}
 {\color{black} In most of our analysis, we will limit our attention considering the system dynamics in the slow-cycling regime $\omega \rightarrow 0$. As we will comment below with reference to some specific examples, the main motivation thereof is that to observe smooth  spectral phase transitions the system evolution must be slow enough}.  Without loss of generality we can assume $G(k)=0$, i.e. $\mathcal{H}_{11}(k)=\mathcal{H}_{22}^*(k)=i \lambda W(k)$, so that the instantaneous eigenenergies of $\mathcal{H}(k)$ read
\begin{equation}
E_{\pm}(k)= \pm \sqrt{R^2(k)-\lambda^2 W^2(k)}
\end{equation}
with $k=\omega t$.
In fact, a non-vanishing value of $G(k)$ can be eliminated from the dynamics after the gauge transformation
\[
\left(
\begin{array}{c}
\psi_1(t) \\
\psi_2(t)
\end{array}
\right)
 \rightarrow 
\left(
\begin{array}{c}
\psi_1(t) \\
\psi_2(t)
\end{array}
\right) \exp \left\{-\frac{i}{ \omega} \int_0^{\omega t} G(k) dk \right\}.
\]
According to Floquet theory, the most general solution to the Schr\"odinger equation (15) is given by
\begin{equation}
 \left(
\begin{array}{c}
\psi_1 (t)  \\
\psi_2 (t)
\end{array}
\right)= \mathcal{U}(t) \exp(-i \mathcal{R} t)  \left(
\begin{array}{c}
\psi_1 (0) \\
\psi_2 (0)
\end{array}
\right)
\end{equation}
where $\mathcal{R}$ is a time-independent $2  \times 2 $ matrix while $\mathcal{U}(t)$ is a time-dependent and periodic $ 2 \times 2 $ matrix, $\mathcal{U}(t+ 2 \pi / \omega)= \mathcal{U} (t)$, with $\mathcal{U}(0)= \mathcal{I}$ (the identity matrix). The exponential of the matrix $\mathcal{R}$ can be expressed in terms of the path-ordered integral 
\[
\exp(-i \mathcal{R} )= \bar{\mathcal{T}} \exp \left[ -i \frac{\omega}{2 \pi}  \int_{0}^{2 \pi / \omega} dt \;\mathcal{H}( \omega t) \right]
\]
where $\bar{\mathcal{T}}$ indicates the  time ordering. 
The two eigenvalues $\mu_{
\pm}=\mu_{\pm}(\lambda)$ of $\mathcal{R}$ are the quasi energies of the time-periodic cycled system. The real parts of the quasi energies are defined apart from integer multiples than $\omega$. Note that for $G(k)=0$ the trace of $\mathcal{H}(k)$ vanishes, so that  $\mu_-=-\mu_+$, i.e. the two quasi energies can be assumed to be opposite one another. A non-vanishing value of $G(k)$ would just lead to a shift of the quasi energies by the amount $(1/2 \pi) \int_0^{2 \pi} dk G(k)$. \\
A natural question arises: akin to the non-cycled $\mathcal{PT}$ symmetric system, is there a spectral phase transition, from real to complex quasi energies, as the non-Hermitian parameter $\lambda$ in the system is increased above a critical value?  To answer this question, let us indicate by $\bar{\lambda}_c$ the minimum value of $\lambda_c(k)$ as $k$ spans the range $0 \leq k \leq 2 \pi$, i.e.
\begin{equation}
\bar{\lambda}_c= \min_{0 \leq k \leq 2 \pi} \lambda_c(k)=\min_{0 \leq k \leq 2 \pi} \left| \frac{R(k)}{W(k)} \right|.
\end{equation}
Intuitively, for a slowly-cycled system one would expect the following scenario: for $\lambda< \bar{\lambda}_c$, the instantaneous eigenenergies $E_{\pm}(k= \omega t)$ of $\mathcal{H}(k=\omega t)$ are real, and thus we expect the quasi energies $\mu_{\pm}$ to remain real as well. On the other hand, for $\lambda> \bar{\lambda}_c$ within the modulation cycle there are time intervals where the instantaneous eigenenergies $E_{\pm}(k= \omega t)$ become complex: in this case we expect the quasi energies to become complex too. Hence, according to such an intuitive picture, we expect a spectral phase transition of the cycled two-level system, from real to complex quasi energies, when the non-Hermitian parameter $\lambda$ is increased above the critical value $\bar{\lambda}_c$. This result is indeed what one observes from a numerical computation of the quasi energies in several examples of cycled two-level $\mathcal{PT}$ symmetric models,  as shown in the next subsection. However, in some other models it turns out that, for a small but non-vanishing oscillation frequency $\omega$, the phase transition is smooth, i.e. imperfect: below the critical value $\bar{\lambda}_c$ the imaginary part of the quasi energy takes a small but non-vanishing value, which scales as $ \sim \omega$, i.e. it exactly vanishes only in the limit $\omega \rightarrow 0$. What is the physical origin of such an imperfect phase transition, which is observed in some models but not in others?\\
 The answer to this question is rooted in the appearance of a complex Berry phase in certain models (but not in others), and can be gained from an adiabatic analysis of the time evolution of the system in the $\omega \rightarrow 0$ limit, which is detailed in the Appendices A and B. In the adiabatic analysis, the slow evolution of the amplitudes of the instantaneous eigenstates $\mathbf{u}_{\pm}(k= \omega t)$ of the Hamiltonian $\mathcal{H}(k =\omega t)$ is governed by the non-Hermitian Berry connection 
\begin{equation}
\mathcal{A}_{n,l}(k)=-i \langle \mathbf{v}_n | \partial_k \mathbf{u}_l \rangle
\end{equation}
($n,l= +,-$), which is defined  in the context of the biorthonormal inner product. The integrals of the diagonal terms of Berry connection, $\mathcal{A}_{+,+}(k)$ and  $\mathcal{A}_{-,-}(k)$, over the interval $0 \leq k \leq 2 \pi$, i.e.
\begin{equation}
\gamma_{B_+}= \int_0^{2 \pi} dk \mathcal{A}_{+,+}(k) \; , \; \; \gamma_{B_-}= \int_0^{2 \pi} dk \mathcal{A}_{-,-}(k)
\end{equation}
are the non-Hermitian Berry phases associated to the two instantaneous eigenstates $\mathbf{u}_{\pm}(k)$.
The explicit form of the Berry connection and Berry phases are derived in Appendix A. In particular, one has
\begin{equation}
\gamma_{B_{\pm}} =  \mp \frac{1}{2} \int_{0}^{2 \pi} dk \frac{d \varphi}{dk} \pm \frac{i}{2} \int_{0}^{2 \pi} dk \frac{d \varphi}{dk} \sinh \psi(k)
\end{equation}
where the function $\psi(k)$ is defined by the relation
 \begin{equation}
 \tanh \psi(k)= \frac{\lambda W(k)}{R(k)}.
 \end{equation}
 Note that the Berry phase vanishes in any $\mathcal{PT}$-symmetric two-level system with $ (d \varphi / dk) \equiv 0$.\\
{\color{black} The adiabatic analysis shows some subtleties and limitations when applied to our model, owing to the appearance of instantaneous EP on the cycle when $\lambda> \bar{\lambda_c}$. Technical details are given in Appendix B}.  The main result of the adiabatic analysis is that, in the limit $\omega \rightarrow 0$ {\color{black} and for $\lambda \neq \bar{\lambda_c}$}, the two quasi energies are given by
\begin{equation}
\mu_{\pm}=  \frac{1}{2 \pi} \int_{0}^{2 \pi} dk E_{\pm} (k) + \frac{\omega}{2 \pi} \gamma_{B_{\pm}}.
\end{equation}
Note that, since $E_-(k)=-E_+(k)$ and $\gamma_{B_{-}}=-\gamma_{B_{+}}$, one has $\mu_-=-\mu_+$, as it should.
The above result provides an approximate form of the quasi energies in the adiabatic limit $\omega \rightarrow 0$ for any strength $\lambda$ of the non-Hermitian parameter far from the critical value $\lambda= \bar{\lambda_c}$, at which the Berry phase term becomes singular and the adiabatic analysis fails; a discussion on this point is given in  the Appendix B.\\
According to Eq.(23), each quasi energy is given by the sum of two terms. The first one is related to the dynamical phase accumulated by the adiabatic eigenstates in one cycle and equals the average of the instantaneous energies $E_{\pm}(k)$ over one cycle. The dynamical phase term is clearly independent of the modulation frequency $\omega$ and is real for $\lambda < \bar{\lambda}_c$, while its imaginary part is non-vanishing for $\lambda > \bar{\lambda}_c$. The second term on the right hand side in Eq.(23) is the non-Hermitian Berry phase contribution. This is a small term which vanishes like $ \sim \omega$ as $\omega \rightarrow 0$.
Interestingly, for $\lambda < \bar{\lambda}_c$ the imaginary part of the quasi energies is provided solely by the Berry phase term, and vanishes as $ \omega \rightarrow 0$. This explains why in the cycled two-level $\mathcal{PT}$-symmetric system with a vanishing Berry phase the spectral phase transition of the quasi energies, at $\lambda=\bar{\lambda}_c$, is sharp (exact), while it becomes smooth (imperfect) when the non-Hermitian Berry phase along the cycle is non-vanishing.

\subsection{Illustrative examples}

The main result of the adiabatic analysis is that the spectral phase transition of the quasi energies in the slow-cycled $\mathcal{PT}$-symmetric two-level system turns out be be imperfect (smooth) whenever the Berry phase in the cycle is complex, which requires the derivative $(d \varphi/dk)$ not to identically vanish. On the other hand, the phase transition is sharp (exact) whenever the Berry phase is real. Here we illustrate and confirm the predictions of the adiabatic analysis by considering three examples of cycled two-level $\mathcal{PT}$-symmetric systems.\\
\\
{1. {\it First example}.} The first example is a simple and exactly-solvable model, corresponding to $G(k)=0$, $W(k)=1$, $R(k)=R_0$, $\varphi(k)=k$, i.e. to the $\mathcal{PT}$-symmetric Hamiltonian
\begin{equation}
\mathcal{H}(k)=
 \left(
\begin{array}{cc}
i \lambda & R_0 \exp(ik) \\
R_0 \exp(-ik) & -i \lambda
\end{array}
\right)
\end{equation}
where $R_0>0$ is a real parameter. Physically, this model describes a two-level system, in which the two states are Hermitian-coupled by an amplitude $R_0$ and a gauge (Peierls) phase $k$ and with gain ($ \lambda$) and loss ($-\lambda$) rates in the two levels. The system displays a $\mathcal{PT}$ symmetry breaking at a critical value $\lambda_c(k)=\bar{\lambda}_c$, independent of $k$, given by
\begin{equation}
\bar{\lambda}_c=R_0.
\end{equation}
In the cycled system with $k= \omega t$,  we expect an imperfect phase transition of quasi energies because $(d \varphi / dk)=1 \neq 0$ and the imaginary part of the Berry phase does not vanish. The quasi energies $\mu_{\pm}$ can be calculated in an exact form, given that the time dependence of the Hamiltonian $\mathcal{H}(k=\omega t)$ can be removed from the dynamics after the gauge transformation
\begin{equation}
\psi_1(t)= \bar{\psi}_1(t) \exp \left( i  \frac{ \omega t}{2} \right) \; , \;  \psi_2(t)= \bar{\psi}_2(t) \exp \left( - i  \frac{\omega t}{2} \right).
\end{equation}
The exact expression of the quasi energies can be readily computed, yielding
\begin{equation}
\mu_{\pm}= \pm \sqrt{R_0^2+ \left( \frac{\omega}{2}+i \lambda \right)^2 } \mp \frac{\omega}{2}
\end{equation}
A typical behavior of the imaginary parts of the quasi energies versus $\lambda$, in the adiabatic limit $ \omega \ll R_0$, is depicted in Fig.1, clearly showing the appearance of an imperfect spectral phase transition near $\lambda=\bar{\lambda}_c$. Note that, when $\lambda$ is not too close to $\bar{\lambda}_c=R_0$, in the adiabatic limit $\omega \rightarrow 0$ we can expand the right hand side of Eq.(27) in power series of $\omega$ and, up to first order in $\omega$, the following approximate expression of quasi energies is obtained
\begin{equation}
\mu_{\pm} \simeq \pm \sqrt{R_0^2-\lambda^2} \mp \frac{\omega}{2}  \pm i \frac{ \lambda \omega}{2  \sqrt{R_0^2-\lambda^2}}.
\end{equation}
It can be readily shown that Eq.(28)  precisely reproduces the result predicted by the adiabatic analysis [Eq.(23)]. In fact, the dynamical phase contribution to the quasi energy is given by
\[
\frac{1}{2 \pi} \int_0^{2 \pi} dk E_{\pm}(k)= \pm \sqrt{R_0^2-\lambda^2} 
\] 

\begin{figure}[!t]
	\centering
	\includegraphics[width=1.0\linewidth]{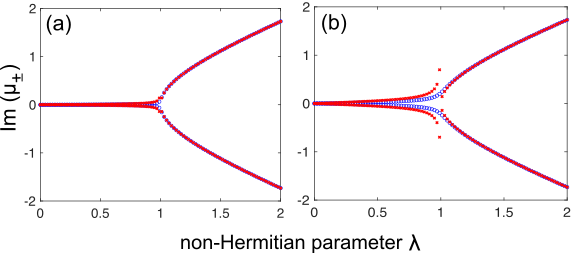}
    \caption{ Behavior of the imaginary part of the quasi energies $\mu_{\pm}$ versus the non-Hermitian parameter $\lambda$ for the cycled $\mathcal{PT}$-symmetric system with Hamiltonian $\mathcal{H}(k)$ given by Eq.(24) for $R_0=1$ and for a modulation frequency (a) $\omega=0.02$, and (b) $\omega=0.1$. Open blue circles and red crosses refer to the exact curves and to the approximate curves obtained from the adiabatic analysis, {\color{black} respectivley}. Note that the spectral phase transition is imperfect around the critical point $\lambda=\bar{\lambda}_c=R_0$, and that near the critical point the adiabatic curves fail to predict the exact behavior of the quasi energies. }
\end{figure}
while the Berry phase contribution reads
\begin{eqnarray}
\frac{\omega}{2 \pi} \gamma_{B_{\pm}} & = &  \mp \frac{ \omega}{4 \pi} \int_{0}^{2 \pi} dk \frac{d \varphi}{dk} \pm \frac{i \omega}{4 \pi} \int_{0}^{2 \pi} dk \frac{d \varphi}{dk} \sinh \psi(k) \nonumber \\
& = &  \mp \frac{\omega}{2} \pm i \frac{\omega}{2} \sinh \psi =\mp \frac{\omega}{2} \pm i \frac{\lambda \omega}{2  \sqrt{R_0^2-\lambda^2}}.
\end{eqnarray}
In deriving Eq.(29), we used the property that $\psi(k)$, defined by the relation ${\rm tanh} \psi(k)= \lambda/R_0$, is independent of $k$ and $(d \varphi/ dk)=1$. Note that the behavior of the imaginary part of the quasi energies, predicted by the adiabatic analysis, well reproduces the exact curves, except near the phase transition point $\lambda=\bar{\lambda}_c$ where the Berry phase contribution displays a singularity.\\
\\
{2. {\it Second example}.} As a second example, let us consider the $\mathcal{PT}$-symmetric two-level Hamiltonian
\begin{equation}
\mathcal{H}(k)=
 \left(
\begin{array}{cc}
i \lambda & t_1+t_2 \cos k \\
t_1+t_2 \cos k & -i \lambda
\end{array}
\right)
\end{equation}
 corresponding to $G(k)=0$, $W(k)=1$, $R(k)=t_1+t_2 \cos k$, and $\varphi(k)=0$, where $t_1$ and $t_2$ are real and positive parameters with $t_1>t_2$. Since $(d \varphi /dk)=0$, the Berry phase vanishes and, according to the adiabatic analysis, when the system is slowly cycled with $k= \omega t$ the spectral phase transition of the quasi energies is sharp (exact) and occurs at the critical value $\bar{\lambda}_c=t_1-t_2$ of the non-Hermitian parameter $\lambda$. The numerical computation of the quasi energies $\mu_{\pm}$ versus $\lambda$, as obtained by a direct numerical integration of the Schr\"odinger equation (15) using an accurate variable-step fourth-order Runge-Kutta method, confirms that the phase transition is sharp and the curves ${\rm Im}(\mu_{\pm}(\lambda))$ are well approximated by the behavior predicted by the adiabatic analysis, as shown in Fig.2.\\
 \\
\begin{figure}[!t]
	\centering
	\includegraphics[width=1.0\linewidth]{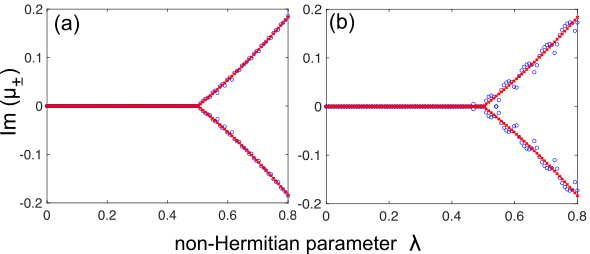}
    \caption{ Behavior of the imaginary part of the quasi energies $\mu_{\pm}$ versus the non-Hermitian parameter $\lambda$ for the cycled $\mathcal{PT}$-symmetric system with Hamiltonian $\mathcal{H}(k)$ given by Eq.(30) for $t_1=1$, $t_2=0.5$ and for a modulation frequency (a) $\omega=0.02$, and (b) $\omega=0.1$. Open blue circles and red crosses refer to the exact curves, obtained from a numerical computation of quasi energies, and to the approximate curves obtained from the adiabatic analysis, {\color{black} respectivley}. Note that the spectral phase transition is sharp around the critical point $\lambda=\bar{\lambda}_c=t_1-t_2=0.5$. }
\end{figure}

{3. {\it Third example}.} As a third example, let us consider the $\mathcal{PT}$-symmetric two-level Hamiltonian
\begin{equation}
\mathcal{H}(k)=
 \left(
\begin{array}{cc}
i \lambda +t_0 \cos k & t_1+t_2 \exp(i k) \\
t_1+t_2 \exp(-ik) & -i \lambda+t_0 \cos k
\end{array}
\right)
\end{equation}
 corresponding to $G(k)=t_0 \cos k$, $W(k)=1$, $R(k)=\sqrt{t_1^2+t_2^2+2 t_1 t_2  \cos k}$, and $\varphi(k)={\rm atan} [t_2 \sin k / (t_1+ t_2 \cos k ]$, where $t_0$, $t_1$ and $t_2$ are real and positive parameters with $t_1 \neq t_2$. Since $(d \varphi /dk) \neq 0$, the imaginary part of the Berry phase does not vanish and, according to the adiabatic analysis, when the system is slowly cycled with $k= \omega t$ the spectral phase transition of the quasi energies is imperfect (smooth). The phase transition occurs at the critical value $\bar{\lambda}_c=|t_2-t_1|$ of the non-Hermitian parameter $\lambda$. The numerical computation of the quasi energies $\mu_{\pm}$ versus $\lambda$, as obtained by a direct numerical integration of the Schr\"odinger equation (15), confirms that the phase transition is imperfect and the curves ${\rm Im}(\mu_{\pm}(\lambda))$ are well approximated by the behavior predicted by the adiabatic analysis for $\lambda \neq \bar{\lambda}_c$, as shown in Fig.3. 
 {\color{black} Note that in the slow-cycling regime [Fig.3(a)] the curves ${\rm Im} (\mu_{\pm})$ versus $\lambda$ display a characteristic knee shape, indicating a smooth spectral phase transition. We remark that the terminology "smooth" phase transition is meaningful in the adiabatic limit $\omega \rightarrow 0$ solely, while when we cycle the system faster, so as $\omega$ becomes comparable to the other characteristic frequencies of the Hamiltonian (such as the separation of adiabatic energies), the knee shape of the curves is continuously spoiled out and there is not any evident sharp transition of the imaginary part of the quasi energies as $\lambda$ is increased; see Fig.3(b).}
 \\
\begin{figure}[!t]
	\centering
	\includegraphics[width=1.0\linewidth]{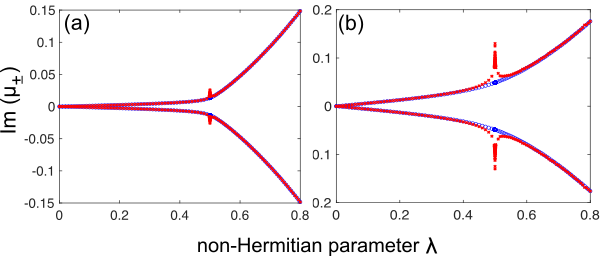}
    \caption{ Behavior of the imaginary part of the quasi energies $\mu_{\pm}$ versus the non-Hermitian parameter $\lambda$ for the cycled $\mathcal{PT}$-symmetric system with Hamiltonian $\mathcal{H}(k)$ given by Eq.(31) for $t_0=0.3$, $t_1=0.5$, $t_2=1$ and for a modulation frequency (a) $\omega=0.02$, and (b) $\omega=0.1$. Open blue circles and red crosses refer to the exact curves, obtained from a numerical computation of quasi energies, and to the approximate curves obtained from the adiabatic analysis. {\color{black} Note that in (a) (slow-cycling limit) the spectral phase transition is smooth around the critical point $\lambda=\bar{\lambda}_c=|t_2-t_1|=0.5$, displaying a characteristic knee shape. In (b) the system is cycled faster and the knee shape of the curves is spoiled out. In both cases  the adiabatic theory fails to the predict the correct behavior of the quasi energies near the critical point.} }
\end{figure}

\section{Wannier-Stark ladder phase transition}
The imperfect spectral phase transition, arising from the complex Berry phase in a slowly-cycled two-level system presented in the previous section, finds an interesting illustrative application to the problem of Wannier-Stark ladder formation in non-Hermitian lattices subjected to a {\color{black} weak} external dc field and the transition from periodic to aperiodic Bloch-Zener oscillations recently observed for some models in Ref.\cite{b81}. In this case the Berry phase is also referred to as the Zak phase \cite{b17}, which is the geometric phase acquired during an adiabatic motion of a Bloch particle across the Brillouin zone.

\subsection{Model}
Let us consider a two-band tight-binding lattice model driven by a dc force $F$. 
In physical space, the temporal evolution of the single-particle state of the system is described by the Schr\"odinger equation
\begin{eqnarray}
i \frac{d a_n}{dt} & = & \sum_{l} \rho_{n-l} a_l+ \sum_{l} \sigma_{n-l}b_l-Fna_n \\
i \frac{d b_n}{dt} & = & \sum_{l} \theta_{n-l} a_l+ \sum_{l} \eta_{n-l}b_l-Fnb_n 
\end{eqnarray}
for the amplitudes $a_n$ and $b_n$ in the two sublattices A and B of the $n$-th unit cell of the crystal. In the above equations, {\color{black} the coefficients 
$\rho_0$ and $\eta_0$ are the on-site energy potentials in the two sublattices A and B, respectively; $\rho_l$  and $\eta_l$ ($l \neq 0$) are the intra-dimer hopping amplitudes; finally,
$\sigma_l$ and $\theta_l$ are the inter-dimer hopping amplitudes}. In the absence of the dc force, i.e. for $F=0$, we can assume $a_n(t)=\psi_1(t) \exp(ik n) $ and $b_n(t) = \psi_2(t) \exp(ikn)$, where $k$ is the Bloch wave number that spans the Brillouin zone $0 \leq k \leq 2 \pi$. In this case, from Eqs.(32) and (33) one obtains
\begin{equation}
i \frac{d}{dt} 
\left(
\begin{array}{c}
\psi_1 \\
\psi_2
\end{array}
\right)= \mathcal{H}(k)
\left(
\begin{array}{c}
\psi_1 \\
\psi_2
\end{array}
\right)
\end{equation}
where the elements of the $ 2 \times 2 $ Bloch Hamiltonian $\mathcal{H}(k)$ are given by
\begin{eqnarray}
\mathcal{H}_{11}(k)= \sum_l \rho_l \exp (-ik l) \\
\mathcal{H}_{12}(k)= \sum_l \sigma_l \exp (-ik l) \\
\mathcal{H}_{21}(k)= \sum_l \theta_l \exp (-ik l) \\
\mathcal{H}_{22}(k)= \sum_l \eta_l \exp (-ik l).
\end{eqnarray}
The Bloch Hamiltonian is $\mathcal{PT}$-symmetric, with $\mathcal{P}=\sigma_x$ and $\mathcal{T}=\mathcal{K}$, provided that 
\begin{equation}
\theta^{*}_{-l}= \sigma_l \; ,\;\; \eta_{-l}^{*}=\rho_l.
\end{equation}
Such conditions ensure that $\mathcal{H}_{22}(k)=\mathcal{H}^*_{11}(k)$ and $\mathcal{H}_{21}(k)=\mathcal{H}^*_{12}(k)$.
In this case, the lattice does not display the non-Hermitian skin effect \cite{skin} and the energy spectrum is absolutely continuous and composed by two energy bands, with the dispersion relation given by Eq.(6). A $\mathcal{PT}$ symmetry breaking phase transition of Bloch bands arises when the non-Hermitian parameter $\lambda$ in the system is increased above the critical value $\bar{\lambda}_c$,  alike in the two-level system discussed in Sec.II. 
\subsection{Wannier-Stark ladders}
When the external dc force is applied, i.e. for $F \neq 0$, the energy spectrum becomes pure point and composed by two Wannier-Stark ladders \cite{WS1,WS2}, with the allowed energies given by 
\begin{equation}
E_l=l F \pm  \Theta 
\end{equation}
where $l=0, \pm 1, \pm 2, \pm 3,..$ and $\Theta$ describes the energy shift of the two ladders. The corresponding eigenstates are normalizable (localized) with a higher-than-exponential localization.\\ 
In an Hermitian lattice, the energy shift $\Theta$ is real and the dynamics in the time domain is generally aperiodic and corresponds to a superposition of Bloch oscillations and Zener tunneling between the two bands \cite{WS1,WS2,WS3,WS4,WS5,WS5bis,WS6,WS7}. The dynamics is 
characterized by two time periods:  The first one, $T_1=2 \pi/F$,
is determined by the mode spacing of each WS ladder and is related to the Bloch oscillation dynamics, whereas the second
one, $T_2= \pi / \Theta$, is
determined by the shift of the two interleaved WS ladders.\\
In a non-Hermitian lattice the energy shift $\Theta$ can become complex and, as we show below, in the small-forcing limit $F \rightarrow 0$ it contains the complex Zak phase of the Bloch Hamiltonian $\mathcal{H}(k)$. More precisely, we will show below that $\Theta$ is the quasi energy $\mu_+$ of the Bloch Hamiltonian $\mathcal{H}(k)$, cycled over the Brillouin zone at a frequency $\omega=F$, i.e. with $k=Ft$. This means that the WS energy spectrum undergoes a phase transition as $\lambda$ is increased above $\bar{\lambda}_c$, from real to complex energies, and the phase transition can be either sharp or smooth, depending on whether the imaginary part of the Zak phase for $\lambda <\bar{\lambda}_c$ is vanishing or not.\\
To calculate the WS energy spectrum $E$, let us assume $a_n(t)=\bar{a}_n \exp(-i Et)$, $b_n(t)=\bar{b}_n \exp(-i Et)$ in Eqs.(32-33), and let us introduce the spectral variables
\begin{eqnarray}
\psi_1(k) & = & \exp(-i E k/F) \sum_n \bar{a}_n \exp(-ikn) \\
\psi_2(k) & = & \exp(-i E k/F) \sum_n \bar{b}_n \exp(-ikn).
\end{eqnarray}
It readily follows that $\psi_{1,2}(k)$ satisfy the Sturm-Liouville problem
\begin{equation}
i F \frac{d}{dk} 
\left(
\begin{array}{c}
\psi_1 \\
\psi_2
\end{array}
\right)= \mathcal{H}(k) 
\left(
\begin{array}{c}
\psi_1 \\
\psi_2
\end{array}
\right)
\end{equation}
on the interval $ 0 \leq k \leq 2 \pi$, with the boundary conditions
\begin{equation}
\psi_{1,2}(2 \pi)=\psi_{1,2}(0) \exp \left( - \frac{2 \pi i E}{F} \right).
\end{equation}
Once the spectral amplitudes $\psi_{1,2}(k)$ and eigenenergies $E$ have been determined, the eigenvectors $ (\bar{a}_n, \bar{b}_n)$, corresponding to the energy $E$, are determined using the inverse relations
\begin{eqnarray}
\bar{a}_n & = & \frac{1}{2 \pi} \int_0^{2 \pi} dk \psi_{1}(k) \exp(ikn+iEk/F) \\
\bar{b}_n & = & \frac{1}{2 \pi} \int_0^{2 \pi} dk \psi_{2}(k) \exp(ikn+iEk/F).
\end{eqnarray} 
Interestingly, after letting $k= \omega t$  Eq.(43) indicates that $\psi_{1,2}(k)$ can be viewed as the amplitudes of a two-level $\mathcal{PT}$-symmetric system, with Hamiltonian $\mathcal{H}(k)$, which is slowly cycled in time at the frequency $\omega=F$. This basically corresponds to the fact that in Bloch space the external force introduces a uniform drift of the quasi-momentum $k$ to span the entire Brillouin zone. The Sturm-Liouville problem, defined by Eqs.(43) and (44), can be solved as follows. Let us indicate by $\boldsymbol\psi_+$ and $\boldsymbol\psi_-$ the eigenvectors of the Floquet matrix $\mathcal{R}$, introduced in Sec.II.B, with eigenvalues (quasi energies) $\mu_{\pm}$. Then Eq.(43) is satisfied by letting either 
\begin{eqnarray}
\left( 
\begin{array}{c}
\psi_1(k) \\
\psi_2(k)
\end{array}
\right)= \mathcal{U} \left( \frac{k}{F}\right) 
  \exp(-i \mathcal{R} k/F) \boldsymbol\psi_+ \nonumber \\ 
=  \exp(-i \mu_+ k/F) \mathcal{U} \left( \frac{k}{R}\right)  \boldsymbol\psi_+ \nonumber
\end{eqnarray}
or 
\begin{eqnarray}
\left( 
\begin{array}{c}
\psi_1(k) \\
\psi_2(k)
\end{array}
\right)= \mathcal{U} \left( \frac{k}{F}\right) 
  \exp(-i \mathcal{R} k/F) \boldsymbol\psi_- \nonumber \\ 
=  \exp(-i \mu_- k/F) \mathcal{U} \left( \frac{k}{F}\right)  \boldsymbol\psi_-. \nonumber
\end{eqnarray}
Since $\mathcal{U}(0)=\mathcal{U}(2 \pi / F)= \mathcal{I}$ (the $ 2 \times 2$ identity matrix), to satisfy the boundary conditions Eq.(44) one should have
\[
\frac{2 \pi}{F} \mu_{\pm}= \frac{2 \pi E}{F}- 2 l \pi
\]
i.e.
\begin{equation}
E=lF+\mu_{\pm}
\end{equation}
where $l=0, \pm 1 , \pm 2, ...$. Equation (47) provides the general form of the Wannier-Stark ladders of allowed energies in terms of the quasi energies $\mu_{\pm}$ of the cycled two-level Bloch Hamiltonian $\mathcal{H}(k)$. It is precisely Eq.(40) with the energy shift parameter $\Theta$ given by  $\Theta=\mu_+$.\\  
{ \color{black} When the external dc force is weak, i.e. in the limit $F \rightarrow 0$, the Bloch wave number $k=Ft$ in the two-level Bloch Hamiltonian $\mathcal{H}(k)$ varies slowly with time, and thus the quasi energies can be approximated by Eq.(23)}. One obtains
\begin{equation}
E=lF+ \frac{1}{2 \pi} \int_{0}^{2 \pi} dk E_{\pm} (k) + \frac{F}{2 \pi} \gamma_{B_{\pm}}
\end{equation}
corresponding to the energy shift 
\begin{equation}
\Theta= \mu_+= \frac{1}{2 \pi} \int_{0}^{2 \pi} dk E_{+} (k) + \frac{F}{2 \pi} \gamma_{B_{+}}.
\end{equation}
\begin{figure}[!t]
	\centering
	\includegraphics[width=1.0\linewidth]{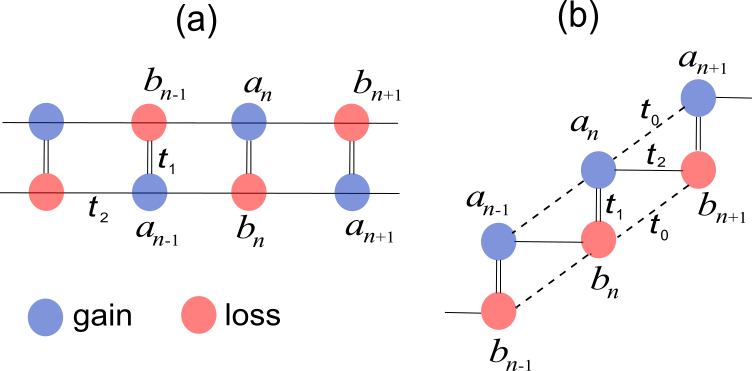}
    \caption{ Schematic of two binary non-Hermitian lattices displaying (a) perfect, and (b) imperfect Wannier-Stark phase transitions under a dc force. $t_0$, $t_1$ and $t_2$ are Hermitian hopping amplitudes. The non-Hermiticity in the system is provided by the gain and loss terms $\pm \lambda$ in the two sublattices A and B. The Bloch Hamiltonian of the two lattices is given by Eq.(30) for model (a), and by Eq.(31) for model (b).}
\end{figure}
Therefore, {\color{black} under a weak external driving} the Wannier-Stark energy spectrum undergoes a phase transition at $\lambda=\bar{\lambda}_c$, which is either sharp or smooth depending on whether the imaginary part of the Zak phase $\gamma_{B{_+}}$ is vanishing or not when $\lambda< \bar{\lambda}_c$. We mention that, contrary to other non-Hermitian lattice models where the spectral ($\mathcal{PT}$ symmetry breaking) phase transition coincides with a localization/delocalization phase transition \cite{uffa1,uffa1bis,uffa2},  in the Wannier-Stark ladder problem the spectral phase transition does not correspond to a localization/delocalization phase transition because the eigenstates of the Wannier-Stark Hamiltonian are always localized, for both $\lambda < \bar{\lambda}_c$ and $\lambda  \geq \bar{\lambda}_c$. This very general result follows from the fact that the spectral amplitudes $\psi_{1,2}(k) \exp(i E k/F)$, with $\psi_{1,2}(k)$ solutions to the Sturm-Liouville problem [Eqs.(43,44)], are periodic and continuously differentiable functions of $k$ and thus their Fourier coefficients $\bar{a}_n$, $\bar{b}_n$ decay as $n \rightarrow \pm \infty$ at least like $ \sim 1/n$, regardless of the value of $\lambda$.\\
In a non-Hermitian lattice below the Wannier-Stark phase transition point, i.e. for $\lambda < \bar{\lambda}_c$, the dynamical signature of a complex Zak phase can be probed looking at the temporal behavior of Bloch-Zener oscillations \cite{b81}. When the imaginary part of the Zak phase is vanishing, the temporal dynamics is rather generally aperiodic and characterized by the two periods $T_1$ and $T_2$, like in an ordinary Hermitian lattice under a dc field: only accidentally the dynamics can be periodic. On the other hand, when the imaginary part of the complex Zak phase does not vanish, after an initial transient the dynamics becomes periodic with period $T_1$. In fact, a rather arbitrary excitation of the system at initial time $t=0$ can be decomposed as a superposition of localized Wannier-Stark eigenstates belonging to the two ladders, and the dynamics at successive times is governed by the interference of such localized eigenstates. The localized Wannier-Stark eigenstates in one ladder, excited by the initial condition, decay in time with a damping rate $ \sim F {\rm{Im}}(  \gamma_{B_+} )$, while the eigenstates in the other ladder are amplified in time with an amplification rate $ \sim F {\rm{Im}} (\gamma_{B_+})$. Therefore, after a transient time of order $ \sim 1/ F {\rm{Im}} (\gamma_{B_+})$ only the Wannier-Stark eigenstates in the former ladder survive and the dynamics become periodic with the period $T_1$ \cite{b81}.

As illustrative examples, let us consider the binary lattices depicted in Figs.4(a) and 4(b). The non-Hermiticity in the lattices is introduced by assuming  energy gain and loss terms $\pm \lambda$ in the two sub lattices A and B. The binary lattice of Fig.4(a) was introduced in a previous work \cite{b74} and its Bloch Hamiltonian $\mathcal{H}(k)$ is given by Eq.(30), previously introduced in Sec.II.C.
Since $( d \varphi / dk) \equiv 0$, the Wannier-Stark ladder phase transition in this model is sharp. The model shown in Fig.4(b) is a non-Hermitian extension of the Rice-Mele model \cite{b54basta1} and its Bloch Hamiltonian is given by Eq.(31). For this model, the spectral phase transition of the Wannier-Stark energies is imperfect. We emphasize that our analysis is very general and could be applied to a generic $\mathcal{PT}$-symmetric binary lattice, also displaying long-range hopping.

  \section{Conclusions and discussion}
 In many classical and quantum systems described by an effective non-Hermitian Hamiltonian, where energy and particles can be exchanged with external reservoirs, the energies of the Hamiltonian are rather generally complex. However, in certain classes of non-Hermitian systems the energy spectrum can remain entirely real in spite of non-Hermiticity.  A paradigmatic example is provided by systems possessing parity-time symmetry, where the energy spectrum remains entirely real in the unbroken $\mathcal{PT}$ phase. When the strength of non-Hermiticity in the system is increased,   a spectral phase transition to complex energies is usually observed, corresponding to the unbroken $\mathcal{PT}$ phase. Such spectral phase transitions are universally sharp. In this work we considered periodically and slowly cycled non-Hermitian models possessing instantaneous $\mathcal{PT}$ symmetry and showed that the phase transition can remain exact (sharp) or become imperfect (smooth) when the strength of non-Hermiticity in the system is increased above a critical value. The imperfect nature of the phase transition in the latter case is universally ascribable to a non-vanishing imaginary part of the complex Berry phase associated to the cyclic adiabatic evolution of the system. This remarkable phenomenon has been illustrated by considering a rather general class of $\mathcal{PT}$-symmetric two-level systems, for which a rigorous adiabatic analysis both below and above the phase transition point has been developed. The results have been applied to describe the spectral phase transitions of the Wannier-Stark ladders in a broad class of $\mathcal{PT}$-symmetric  two-band non-Hermitian lattices subjected to an external dc field, {\color{black} however our analysis is expected to hold for more general multi-band systems. In fact, under the adiabatic conditions and assuming no state flip after one adiabatic cycle, the form of quasi energies can be given in terms of dynamic and geometric (Zak) phases, and the complex or real nature of the latter defines the smooth or sharp nature of the spectral phase transitions in the slow-cycling regime.}
 Our results provide fresh and novel insights into phase transitions of open quantum or classical systems, providing important examples of smooth phase transitions in non-Hermitian physics and unraveling the main role played by the non-Hermitian Berry phase.

\acknowledgments
S.L. acknowledges the Spanish State Research Agency, through the Severo Ochoa
and Maria de Maeztu Program for Centers and Units of Excellence in R\&D (Grant No. MDM-2017-0711). L.F. acknowledges the support from National Science Foundation (NSF) (ECCS-1846766).

\appendix
\section{Berry connection and Berry phase}
For the cycled two-level $\mathcal{PT}$ symmetric model considered in Sec.II.A, the elements of the $2 \times 2$ matrix of the non-Hermitian Berry connection are given in terms of the biorthogonal product as 
  \begin{equation}
  \mathcal{A}_{n,l}= -i \langle \mathbf{v}_n | \partial_k \mathbf{u}_l \rangle 
  \end{equation}
  where $n,l$ take the values $+$ or $-$. Using Eqs.(7,8) and (10,11) given in the main text, the explicit form of the Berry connection can be readily calculated and read
  \begin{eqnarray}
  \mathcal{A}_{+,+} & = & - \frac{d \varphi}{dk} \sin^2 \left( \frac{\theta}{2} \right) \\
  \mathcal{A}_{+,-} & = &  - \frac{1}{2} i \frac{d \theta}{dk} + \frac{1}{2} \frac {d \varphi}{dk} \sin \theta \\
  \mathcal{A}_{-,+} & = &  \frac{1}{2} i \frac{d \theta}{dk} + \frac{1}{2} \frac {d \varphi}{dk} \sin \theta \\
  \mathcal{A}_{-,-} & = & - \frac{d \varphi}{dk} \cos^2 \left( \frac{\theta}{2} \right).
  \end{eqnarray}
The Berry phases associated to the two adiabatically-evolving eigenstates $\mathbf{u}_{\pm}(k)$ are given by
\begin{eqnarray}
\gamma_{B_{+}} &  \equiv & \int_0^{2 \pi} dk \mathcal{A}_{+,+}=- \int_0^{2 \pi} dk \frac{d \varphi}{dk} \sin^2 \left( \frac{\theta}{2} \right) \;\;\;\;  \\
\gamma_{B_{-}} &  \equiv & \int_0^{2 \pi} dk \mathcal{A}_{-,-}=- \int_0^{2 \pi} dk \frac{d \varphi}{dk} \cos^2 \left( \frac{\theta}{2} \right). \;\;\;\;
\end{eqnarray}
 From Eqs.(A6) and (A7) it readily follows that $\gamma_{B_{+}}+\gamma_{B_{-}}=- \int_0^{2 \pi} dk (d \varphi / dk)$ is either zero or an integer multiple than $ 2 \pi$. Since the Berry phase is defined apart from integer multiples than $2 \pi$,  we can thus write 
\begin{eqnarray}
\gamma_{B_{+}} & = & -\gamma_{B_{-}}= \\
& = & -\frac{1}{2} \int_0^{2 \pi} dk \frac{d \varphi}{dk}+\frac{1}{2} \int_0^{2 \pi} dk \cos \theta \frac{d \varphi}{dk}. \nonumber
\end{eqnarray}
The complex angle $\theta=\theta(k)$ is defined by Eq.(9) given in the main text, i.e.
\begin{equation}
\tan \theta(k) = \frac{R(k)}{i \lambda W(k)}
\end{equation}
 which can be solved by letting 
 \begin{equation}
 \theta(k)= \pi /2-i \psi(k), 
 \end{equation}
 where the function $\psi(k)$ is given by
 \begin{equation}
 \tanh \psi(k)= \frac{\lambda W(k)}{R(k)}.
 \end{equation}
 Using Eqs.(A8) and (A10), one finally obtains
\begin{equation}
\gamma_{B_{\pm}} =  \mp \frac{1}{2} \int_{0}^{2 \pi} dk \frac{d \varphi}{dk} \pm \frac{i}{2} \int_{0}^{2 \pi} dk \frac{d \varphi}{dk} \sinh \psi(k)
\end{equation}
The above expression of the Berry phase is formally valid for any value of the non-Hermitian parameter, except for $\lambda= \bar{\lambda}_c$.
Note that for $\lambda< \bar{\lambda}_c$ one has $| \lambda W(k) /R(k)|<1$ and thus the function $\psi(k)$ is real over the entire interval $ 0 \leq k \leq 2 \pi$. In this case Eq.(A12) shows that the real part of the Berry phase is quantized and can take only the two values $0$ or $ \pi$ (mod. $ 2 \pi$), whereas the imaginary part of the Berry phase is not quantized and vanishes whenever $(d \varphi/ dk) \equiv 0$. 

\section{Adiabatic analysis}
In this Appendix we derive an analytical expression of the quasi energies $\mu_{\pm}$ of the cycled two-level $\mathcal{PT}$ symmetric system, considered in Sec.II.B of the main text, in the adiabatic limit of slow cycling $\omega \rightarrow 0$. It should be mentioned that special attention is required when using adiabatic methods to slowly-evolving non-Hermitian systems because: (i) Owing to possible nontrivial topologies of the energy curves in complex plane, even in absence of eigenvalue degeneracies it could happen that an adiabatically evolving eigenstate, after one cycle, does not come back to its initial state because of energy and eigenvector flipping \cite{b58,b59}; (ii) Even in the slow cycling regime the adiabatic approximation can easily break down when the instantaneous eigenenergies are complex \cite{bf1,bf2,bf3}, and the adiabatic approximation can be safely applied only to the most dominant eigenstate of the system.\\
After letting $k= \omega t$, the Schr\"odinger equation (15) reads
\begin{equation}
i \omega \frac{d}{dk} 
\left(
\begin{array}{c}
\psi_1 \\
\psi_2
\end{array}
\right) =
\mathcal{H}(k)  \left(
\begin{array}{c}
\psi_1 \\
\psi_2
\end{array}
\right).
\end{equation}
 To perform the adiabatic analysis, let us distinguish two cases.\\
 {\it First case: $\lambda < \bar{\lambda}_c$.} In this case the two energy curves $E_{\pm}(k)$, as $k$ spans the interval $(0, 2 \pi)$, are straight and non-intersecting segments on the real energy axis; see Fig.5(a). Therefore, the energy curves are line gapped and there is not any eigenvalue/eigenstate flip after one cycle. From the point of view of the adiabatic analysis, the system thus behaves like an Hermitian one, even though the Hamiltonian is not Hermitian. We then expand the state vector $(\psi_1(k), \psi_2(k) )^T$ as a superposition of the instantaneous eigenstates $\mathbf{u
}_+(k)$ and $\mathbf{u}_-(k)$ of $\mathcal{H}(k)$, i.e. let us set
\begin{eqnarray}
\left(
\begin{array}{c}
\psi_1(k) \\
\psi_2(k)
\end{array}
\right) & = &  a_+(k) \mathbf{u}_+(k) \exp \left\{ -\frac{i}{\omega} \int_0^k d \xi E_+ (\xi) \right\} \nonumber \\
& + & a_-(k) \mathbf{u}_-(k) \exp \left\{ - \frac{i}{\omega} \int_0^k d \xi E_-(\xi) \right\} \;\;\;\;\;\;
\end{eqnarray}
  where $a_+(k)$ and $a_-(k)$ are the adiabatic amplitudes and
  \begin{equation}
  E_{\pm}(k)= \pm \sqrt{R^2(k)-\lambda^2 W^2(k)}
  \end{equation}
  are the instantaneous eigenenergies. The evolution equations of the amplitudes $a_{\pm}(k)$ are readily obtained by substitution of the Ansatz (B2) into Eq.(B1) and taking the scalar product of the equation so obtained by $ \langle \mathbf{v}_+|$ and $ \langle \mathbf{v}_-|$. Using the biorthogonal conditions (12), one obtains
  \begin{eqnarray}
  i \frac{ da_+}{dk}   & = &    \mathcal{A}_{+,+} a_+ + \\
   & + &   \mathcal{A}_{+,-} a_- \exp \left\{ \frac{i}{\omega} \int_0^k d \xi \left[ E_+ (\xi)-E_- (\xi) \right]  \right\}  \nonumber \\
  i \frac{ da_-}{dk}  & = &  \mathcal{A}_{-,-} a_- +\\
  & + &  \mathcal{A}_{-,+} a_+ \exp \left\{ -\frac{i}{\omega} \int_0^k d \xi \left[ E_+ (\xi)-E_- (\xi) \right]  \right\} \nonumber
  \end{eqnarray}
  where $ \mathcal{A}_{n,l}$ ($n,l=+,-)$ is the non-Hermitian Berry connection, given by Eqs.(A1-A5).
In the adiabatic limit $\omega \rightarrow 0$, since the energy difference $E_+(k)-E_-(k)$ is entirely real and non-vanishing over the interval $0 \leq k \leq 2 \pi$, the rapidly oscillating terms on the right hand sides of Eqs.(B4) and (B5) do not induce on average transitions between the two adiabatic amplitudes and can be disregarded (rotating-wave approximation). Hence one obtains
\begin{equation}
a_{\pm}(2 \pi) \simeq a_{\pm}(0) \exp(-i \gamma_{B_{\pm}})
\end{equation}
where $\gamma_{B_{\pm}}$ are the Berry phases associated to the two adiabatically-evolving eigenstates $\mathbf{u}_{\pm}(k)$. The explicit form of the Berry phases is given by Eq.(A12).\\
The quasi energies $\mu_{\pm}$ are the eigenvalues of the matrix $\mathcal{R}$, which is obtained from the condition [Eq.(17) in the main text with $t= 2 \pi / \omega$]
\begin{equation}
 \left(
\begin{array}{c}
\psi_1 ( 2 \pi / \omega)  \\
\psi_2 ( 2 \pi / \omega)
\end{array}
\right)= \exp(-2 \pi i \mathcal{R} / \omega)  \left(
\begin{array}{c}
\psi_1 (0) \\
\psi_2 (0)
\end{array}
\right).
\end{equation}
From Eqs.(B2) and (B5) it readily follows that $\mathbf{u}_{\pm}(0)$ are the eigenvectors of $\mathcal{R}$ with corresponding quasi energies given by
\begin{equation}
\mu_{\pm}=  \frac{1}{2 \pi} \int_{0}^{2 \pi} dk E_{\pm} (k) + \frac{\omega}{2 \pi} \gamma_{B_{\pm}}.
\end{equation}
Note that, since $E_{-}(k)=-E_+(k)$ and $\gamma_{B_{-}}=-\gamma_{B_{+}}$, the two quasi energies are opposite one another, i.e. $\mu_-=-\mu_+$, as it should be whenever $G(k)=0$. Note also that each quasi energy is given by the sum of two terms. The first term on the right hand side of Eq.(B8) is the usual dynamical phase term that one would obtain by a standard WKB analysis neglecting the Berry phase \cite{b77,b81}, whereas the second term on the right hand side of Eq.(B8) is the Berry phase contribution. While the dynamical phase term is always real and independent of the modulation frequency $\omega$, the Berry phase contribution vanishes as $ \omega \rightarrow 0$ and can display a nonvanishing imaginary part in models where $(d \varphi /dk) \neq 0$. Therefore, we may conclude that for $\lambda < \bar{\lambda}_c$ the imaginary part of the quasi energies, as predicted by the adiabatic analysis, reads
\begin{equation}
{\rm \Im} ( \mu_{\pm} ) = \frac{\omega}{2 \pi} {\rm{Im}} ( \gamma_{B_{\pm}} )=  \pm \frac{\omega}{4 \pi}  \int_{0}^{2 \pi} dk \frac{d \varphi}{dk} \sinh \psi(k)
\end{equation}
where the real function $\psi(k)$ is defined by Eq.(A11).\\
\begin{figure}[!t]
	\centering
	\includegraphics[width=1.0\linewidth]{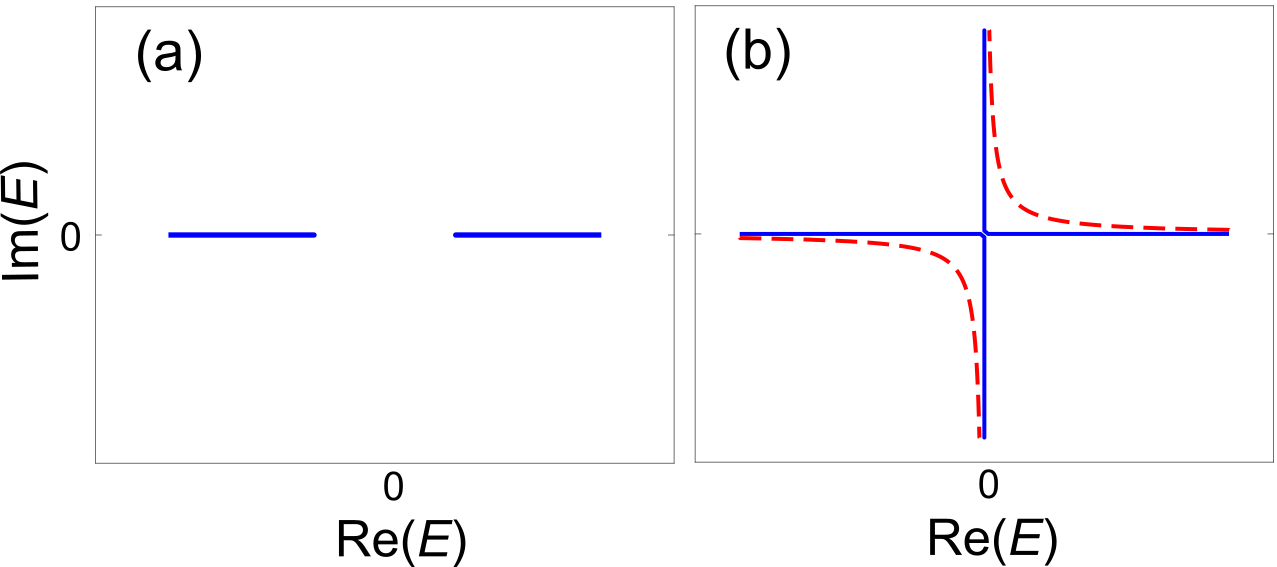}
    \caption{ Schematic behavior of the energy curves $E_{\pm}(k)$ of the $\mathcal{PT}$-symmetric two-level Hamiltonian $\mathcal{H}(k)$ in complex energy plane as $k$ spans the interval $0 \leq k \leq 2 \pi$ (solid lines). In (a) $\lambda < \bar{\lambda}_c$, the two energy curves lie on the real energy axis and are line gapped. In (b) $\lambda> \bar{\lambda}_c$ and the two energy curves cross at $E=0$ (instantaneous exceptional point) at the critical values $k=k_c$ such that $\lambda W(k_c)= \pm R(k_c)$. The dashed curves in (b) show the behavior of the energy curves for the modified Hamiltonian $\mathcal{H}_{\epsilon}(k)$, which avoids energy crossing and exceptional points.}
\end{figure}
\\
{\it Second case: $\lambda > \bar{\lambda}_c$.} In this case the two energy curves $E_{\pm}(k)$, as $k$ spans the interval $(0, 2 \pi)$, may touch one another at $E=0$, as shown by the solid curves in Fig.5(b). The crossing occurs when $k$ equals the critical values $k=k_c$ such that $\lambda W(k_c)= \pm R(k_c)$. At such points, the instantaneous Hamiltonian $\mathcal{H}(k_c)$ is not diagonalizable and displays an exceptional point. Eventually, if $W(k)$ does not vanish in the entire range $(0, 2 \pi)$, at large values of $\lambda$ the two energy curves can become separated and fully lie on the imaginary axis.\\ 
The occurrence of the instantaneous exceptional points and energy curve touching  at $k=k_c$ during the cycle when $\lambda> \bar{\lambda}_c$ makes it formally invalid the adiabatic analysis discussed in the previous case. To overcome such a limitation, we slightly modify the Hamiltonian of the system, from $\mathcal{H}(k)$ to $\mathcal{H}_{\epsilon}(k)$, by letting
\begin{equation}
\left( \mathcal{H}_{\epsilon} (k)  \right)_{11}=i \lambda W(k)+ \epsilon \; , \; \; \left( \mathcal{H}_{\epsilon} (k) \right)_{22}=-\left( \mathcal{H}_{\epsilon} (k)  \right)_{11}
\end{equation}
where $\epsilon>0$ is a small real parameter. For $\epsilon=0$ we recover the original Hamiltonian $\mathcal{H}(k)$. The instantaneous eigenenergies of $\mathcal{H}_{\epsilon}(k)$ read
\begin{equation}
E_{\epsilon \; \pm}= \pm \sqrt{R^2(k)-(\lambda W(k)-i \epsilon)^2}.
\end{equation}
A non-vanishing (albeit small) value of $\epsilon$ breaks exact $\mathcal{PT}$ symmetry and avoids the energy curve touching and the appearance of the instantaneous exceptional points during the adiabatic cycle, as shown by the dashed curves in Fig.5(b). Since the two energy curves are now line gapped and there are not exceptional points along the cycle, we can again expand the state vector of the system as a superposition of the instantaneous eigenstates of  $\mathcal{H}_{\epsilon}(k)$ with adiabatic amplitudes $a_{\pm}(k)$, which evolve according to Eqs.(B4) and (B5) (these are \emph{exact} equations). The main difference is that the Berry connection and instantaneous eigenenergies entering in such equations are now those of the modified Hamiltonian  $\mathcal{H}_{\epsilon}(k)$ rather than  $\mathcal{H}(k)$. 
The expressions of the instantaneous (right and left) eigenstates  of $\mathcal{H}_{\epsilon}(k)$, and thus of Berry connection and Berry phases, are formally the same as those of $\mathcal{H}(k)$, with the complex angle $\theta=\theta(k)$ now defined by the relation
\begin{equation}
\tan \theta(k) = \frac{R(k)}{i \lambda W(k)+ \epsilon}.
\end{equation}
It should be noted that, as $ \epsilon \rightarrow 0$ the imaginary part of $\theta(k)$ diverges at the critical values  $k=k_c$, however for a linear crossing, such that $ \lambda W'(k_c) \neq  \pm R'(k_c)$, the singularity of $\cos \theta(k)$ near $k=k_c$ is of the type $ \cos \theta(k) \sim 1 / \sqrt{k-k_c}$ and thus integrable, leading to a finite value of the Berry phase according to Eq.(A8).\\ 
To calculate the quasi energies, we exploit the fact that $\mu_{-}=-\mu_+$, so that we can compute the quasi energy of the dominant adiabatic eigenstate of the system, i.e. with the largest imaginary part of instantaneous energy [corresponding to the dashed curve in the first quadrant of Fig.5(b)]. For such a state we can in fact safely apply the adiabatic approximation, avoiding the problem of adiabaticity breakdown that could arise for the non-dominant eigenstate \cite{bf1,bf2}. For example, assuming that $\mathbf{u}_+(k)$ is the dominant instantaneous eigenstate, i.e. with ${\rm Im}(E_+(k)) \geq 0$, we can safely apply the rotating-wave approximation to the second term on the right hand side of Eq.(B4), thus obtaining $a_+(2 \pi) \simeq a_+(0) \exp(-i \gamma_{B_{+}})$. Proceeding as in the previous case, in the adiabatic limit one then obtains the following expression of the quasi energy $\mu_+$, associated to the dominant adiabatic eigenstate
\begin{equation}
\mu_{+}=  \frac{1}{2 \pi} \int_{0}^{2 \pi} dk E_{+} (k) + \frac{\omega}{2 \pi} \gamma_{B_{+}}.
\end{equation}
The other quasi energy is then given by $\mu_-=-\mu_+$. This result formally coincides  with the one obtained in the case $\lambda < \bar{\lambda}_c$ [see Eq.(B8)], however in Eq.(B13) the Berry phase term should be obtained from the modified angle $\theta$, given by Eq.(B12), and then taking the limit $\epsilon \rightarrow 0$. The main difference in the $\lambda > \bar{\lambda}_c$ case is that the dynamical phase contribution to the quasi energy has a non-vanishing imaginary part, which dominates over the imaginary contribution of the Berry phase term in the adiabatic ($ \omega \rightarrow 0$) limit.\\
 Finally, we mention that at the phase transition point $\lambda=\bar{\lambda}_c$ the crossing of the exceptional point during the oscillation cycle, at $k=k_c$, is quadratic rather than linear, i.e. one has $\lambda W(k_c)= \pm R(k_c)$ and $\lambda W{'}(k_c)= \pm R{'}(k_c)$. In this case, as $\epsilon \rightarrow 0$  the singularity of $\cos \theta(k)$ near $k=k_c$ is of the type $ \cos \theta(k) \sim 1 / (k-k_c)$ and thus it is not integrable,  leading to a diverging value of the Berry phase according to Eq.(A8). Therefore, the adiabatic analysis fails to predict the correct values of the quasi energies as $\lambda$ approaches the critical value $\bar{\lambda}_c$, either from below or from above. Such a failure is clearly illustrated in the exactly-solvable model with Hamiltonian $\mathcal{H}(k)$ given by Eq.(24), discussed in the main text [see specifically Eq.(29) and Fig.1(b)].


\end{document}